\pdfoutput=1

\documentclass[11pt]{article}

\usepackage{acl}

\usepackage{times}
\usepackage{latexsym}

\usepackage[T1]{fontenc}

\usepackage[utf8]{inputenc}

\usepackage{microtype}

\usepackage{inconsolata}

\usepackage{enumerate}
\usepackage{enumitem}
\usepackage[group-separator={,}]{siunitx}
\usepackage{graphicx}
\usepackage{xspace}
\usepackage{amsfonts}
\usepackage{threeparttable}
\usepackage{booktabs}
\usepackage{multirow}
\usepackage{subfigure}
\usepackage{makecell}

\usepackage{bbm}

\newtheorem{problem}{Problem}

\newcommand{\hide}[1]{}
\newcommand{\vpara}[1]{\vspace{0.07in}\noindent\textbf{#1 }}
\newcommand{\beq}[1]{\vspace{-0.1in}\begin{equation}#1\end{equation}\vspace{-0.1in}}

\newcommand{\refsource}{\textit{ref-source}\xspace}
\newcommand{\refsources}{\textit{ref-sources}\xspace}
\newcommand{\pstbench}{PST-Bench\xspace}

\usepackage{amssymb}

\newlist{todolist}{itemize}{2}
\setlist[todolist]{label=$\square$}
\usepackage{pifont}
\newcommand{\cmark}{\ding{51}}%
\newcommand{\xmark}{\ding{55}}%
\newcommand{\done}{\rlap{$\square$}{\raisebox{2pt}{\large\hspace{1pt}\cmark}}%
\hspace{-2.5pt}}
\newcommand{\wontfix}{\rlap{$\square$}{\large\hspace{1pt}\xmark}}

\title{\pstbench: Tracing and Benchmarking the Source of Publications}

\author{Fanjin Zhang$^{1,2}$\thanks{\ Partial work was done when Fanjin worked at Zhipu AI.}, Kun Cao$^2$, Yukuo Cen$^2$, Jifan Yu$^1$, Da Yin$^2$, Jie Tang$^1$\\
$^1$ Department of Computer Science and Technology, Tsinghua University, Beijing, China\\
 $^2$ Zhipu AI, Beijing, China \\
 \{fanjinz, jietang\}@tsinghua.edu.cn, bzsy2476203449@gmail.com\\
 \{yukuo.cen, da.yin\}@aminer.cn, yujf21@mails.tsinghua.edu.cn
}

\begin{document}
\maketitle
\begin{abstract}
Tracing the source of research papers is a fundamental yet challenging task for researchers.
The billion-scale citation relations between papers hinder researchers from understanding the evolution of science efficiently.
To date, there is still a lack of an accurate and scalable dataset constructed by professional researchers 
to identify the direct source of their studied papers,
based on which automatic algorithms can be developed to expand the evolutionary knowledge of science.
In this paper, we study the problem of paper source tracing (PST)
and construct a high-quality and ever-increasing dataset \pstbench in computer science.
Based on \pstbench,
we reveal several intriguing discoveries, such as the differing evolution patterns across various topics.
An exploration of various methods underscores the hardness of \pstbench,
pinpointing potential directions on this topic.
The dataset and codes have been available\footnote{\url{https://github.com/THUDM/paper-source-trace}}.
\end{abstract}

\section{Introduction}
\label{sec:intro}

\begin{figure}[t]
	\centering
	\includegraphics[width=6cm]{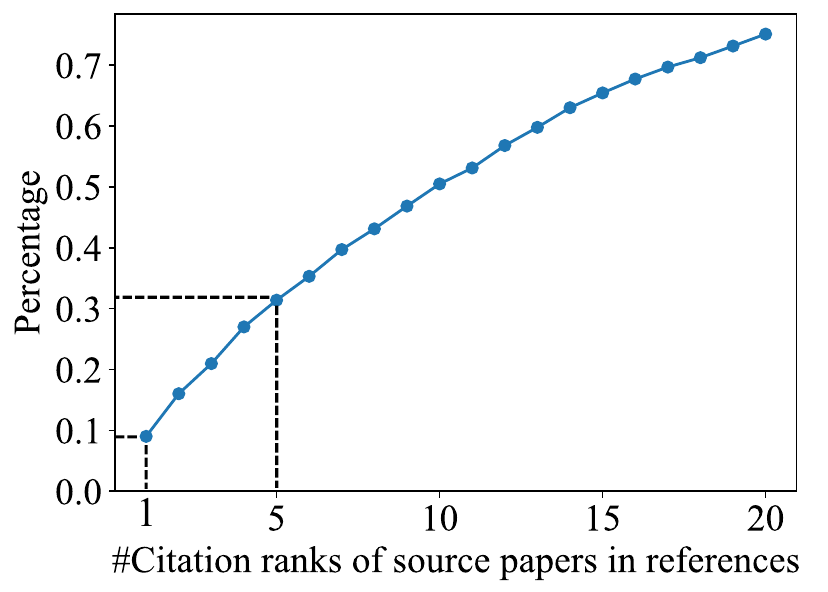}
    \caption{The cumulative distribution function (CDF) of the references' citation ranks for source papers.}
	\label{fig:cdf_citation_rank}
\end{figure}

Comprehending the patterns of scientific evolution,
such as the trends of topics and the flow of ideas,
are critical 
for funding agencies in policy development and for researchers in knowledge discovery
~\cite{fortunato2018science}.
The trajectory of scientific evolution can be discerned through citation relationships.
However, 
a notable gap persists between
the large-scale and semantically rich citation relations~\cite{zhang2019oag,tang2009topic}
and the backbone structure of scientific evolution.
To reveal the essence of scientific development, 
how can we simplify the citation graph to 
trace the source of publications and 
uncover the relationships of inspiration between papers?
One might intuitively consider the most cited references of each paper as the sources, 
discarding other citation relations.
However, this is not the case.
Based on around $1{\small,}500$ computer science papers
and their professionally annotated source papers,
we visualize the cumulative distribution function (CDF) 
of the references' citation ranks of source papers 
in Figure \ref{fig:cdf_citation_rank}.
It shows that if we consider the most cited reference as the source paper,
the accuracy is less than $10\%$.
Further, nearly $70\%$ of papers 
have source papers that are not among the top-$5$ cited references.
This challenges the intuition that the citation number is 
the primary indicator to identify the source of publications.
For example, Random Forest~\cite{breiman2001random}, 
Scikit-learn~\cite{pedregosa2011scikit}, and ImageNet~\cite{deng2009imagenet} 
are among the most cited papers.
However, they are not frequently regarded as the direct sources of annotated papers 
as these works are popular and classic methods/tools/benchmarks. 
In contrast, the TAGE branch predictor~\cite{seznec2006case},
a performance-critical component in modern CPUs,
receives less than $20$ citations per year on average,
but its inspired variants are applied to most high-end ARM processors~\cite{pellegrini2021arm} 
and AMD Zen processors~\cite{suggs2020amd}.

Tracing the source of publications is a challenging issue  
that remains under-explored. 
~\citet{valenzuela2015identifying} classify citing relationships into incidental and important citations
and propose a feature-engineering approach to predict important citations.
However, their dataset only comprises less than $100$ annotated important citing pairs.
Given the high level of expertise required for annotation, many related works employ automated methods to generate datasets.
Algorithm Roadmap~\cite{zha2019mining} 
applies weak supervision in the citation contexts to generate datasets and
extract comparative algorithms from texts.
Further, MRT~\cite{yin2021mrt} is an unsupervised framework designed to 
generate fine-grained annotated evolution roadmaps for specific publications
by utilizing text embeddings and node embeddings on citation graphs.
MRT assesses the generated important scores between papers and references 
based on user clicks on the generated roadmap,
which may suffer from the sparsity and bias of user clicks.
Consequently, relevant resources suffer from 
\textit{small scale}, \textit{lack of diversity due to machine generation},
or \textit{absence of professional annotation}.

\hide{
Until now, to grasp the ins and outs of technological development from vast literature,
it becomes indispensable to trace the source of papers.
Otherwise, researchers may find themselves inundated with a multitude of papers and a vast array of references.
However, this problem presents
the following challenges:
(1) How to formally define the source of a paper?
(2) How to construct a high-quality and ever-increasing dataset for paper source tracing?
(3) What are the underlying patterns behind the paper source tracing graph?
(4) Is it feasible to automatically trace the source of papers?
}

\vpara{Present Work.}
For this purpose, 
in this study, we first formally define the problem of paper source tracing (\textbf{PST}) 
and introduce \textbf{PST-Bench}, a professionally annotated PST dataset
comprising $1{\small,}576$ computer science papers and \num{55014} associated references,
supplemented by additional $4{\small,}800$ papers and their rule-generated source papers. 
Each target paper within this dataset has been meticulously annotated with its source papers.
We devise a new data annotation strategy via an online paper reading group
to ensure high-quality and ever-increasing professional annotations.
Second, we perform a comprehensive analysis of this dataset,
examining aspects such as the year gap and cross-venue influence 
between papers on different topics and their source papers,
uncovering several interesting patterns.
Lastly, we investigate the potential for automatically tracing the source of papers.
To summarize, our contributions are as follows.

\begin{itemize}[leftmargin=*]
    \item We establish an accurate, diverse, and continually expanding paper source tracing dataset \textbf{PST-Bench}.
    To achieve this, we develop a novel strategy that leverages a reading group of graduate students 
    to share papers and mark the sources of papers accurately and regularly.
    \item We perform in-depth analyses of the PST graph, 
    revealing several intriguing discoveries.
    For instance, 
    papers in high performance computing (HPC)
    tend to draw inspiration from less-cited papers than AI papers,
    even though the former are inclined to be influenced by older papers.
    
    \item We explore a variety of methods to automatically trace the source of papers,
    including statistical methods, graph-based methods, and pre-trained language model (PLM) based methods. 
    Experiments indicate that PLMs exhibit the potential for addressing the PST problem.
    However, the best result of automatic methods is still far from satisfactory,
    leaving much room for future research.
\end{itemize}

\pstbench can be used for various research topics, such as
understanding scientific evolution, 
studying automatic paper source tracing,
and measuring paper impact,
aiming to boost
innovation through analogy mining and thinking ultimately.

\section{Problem Definition}

In this section, we formally define the problem of paper source tracing (PST).

\begin{problem}
    \textbf{Paper Source Tracing (PST)}.
    Given a target paper $p$ along with its full text,
    the objective is to identify the most important references,
    termed as 
    ``\refsources'', 
    that have significantly contributed to the ideas or methods presented in the paper.
    For each reference within the paper $p$,
    an important score ranging from $0$ to $1$ should be assigned,
    indicating the degree of influence each reference has exerted on the paper.
    For each paper $p$, the predictive output is denoted as $S_p$.
\end{problem}

Note that a paper may draw inspiration from one or more \textit{``ref-sources''}. 
The determination of whether a reference qualifies as a \textit{``ref-source''} is based on one of the following criteria:

\begin{itemize}[leftmargin=*]
    \item Does the main idea of paper $p$ draw inspiration from the reference?
    \item Is the fundamental methodology of paper $p$ derived from the reference?
\end{itemize}

Namely,
is the reference indispensable to paper $p$?
Without the contributions of the reference, 
would the completion of paper $p$ be impossible?
It's vital to clarify that if paper $p_c$ cites both papers $p_a$ and $p_b$, 
with $p_a$ serving as a \textit{ref-source} for $p_b$ 
and $p_b$ in turn serving as a \textit{ref-source} for $p_c$. 
In this case, $p_a$ does not become a \textit{ref-source} for $p_c$, 
even if $p_c$ cites $p_a$. 
Our focus is solely on identifying \textit{ref-sources} 
that \textbf{directly} inspire paper $p$.

\section{Building the \pstbench}
\label{sec:dataset}

Considering the specialized knowledge necessary for tracing the sources of academic papers, 
we engaged dozens of computer science graduate students to identify the sources of English papers within their respective fields of expertise. 

\begin{figure}[t]
	\centering
	\includegraphics[width=.5\textwidth]{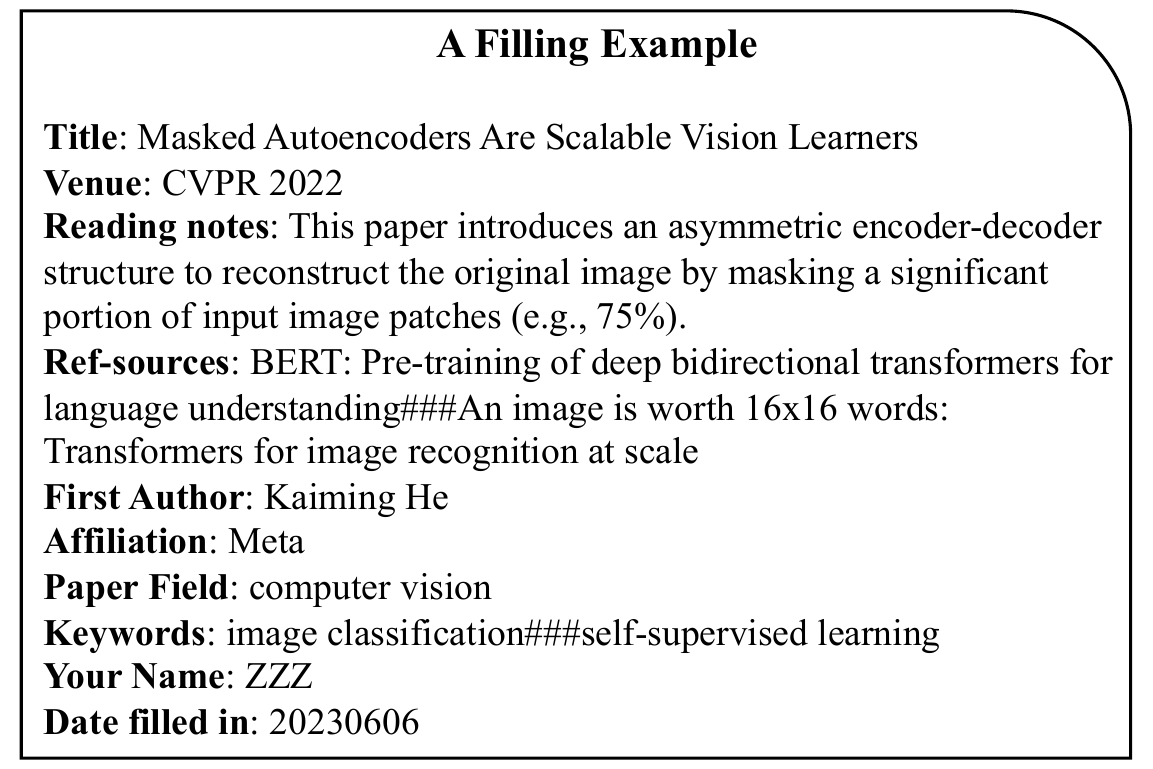}
    \caption{A filling example.
    Multiple items are separated by ``\#\#\#''
    in the fields of \textit{ref-sources} and keywords.}
	\label{fig:reading_group_example}
\end{figure}

Our data collection methodology is bifurcated into two approaches. 
The first approach involves each student marking the papers they had previously read, 
averaging around $20$ papers per individual. 
To ensure a consistent influx of high-quality labeled data, 
the second approach requires each student to read and mark two new papers 
every week. 
This is conducted in the format of an \textit{online paper reading group}, 
where students identify the source papers of the ones they read recently.
A data collection example is shown in Figure~\ref{fig:reading_group_example}.
More specifics about data collection can be found in Section \ref{sec:app:data_collection}.

After gathering and preprocessing the data, 
we obtain a total of $1{\small,}576$ labeled computer science papers. 
The dataset is then partitioned based on their publication year, 
with $788$ papers allocated for training, 
$394$ for validation, and the remaining $394$ set aside for testing.

Furthermore, we additionally generate a supplementary dataset by extracting references that appear near signal words like ``motivated by'' and ``inspired by'' as source papers,
resulting in $4{\small,}800$ papers with their rule-generated source papers.

\vpara{Quantity control \& quality control.}
We devise several strategies to ensure a 
steady and high-quality
growth of the dataset.
First, each student only needs to read and mark two new papers every week,
avoiding the attacks of perfunctory annotations to some extent.
Second, we offer additional accumulated rewards to students 
once they have read and marked a certain number of papers (e.g., $20$) 
and remove students who have not marked any papers for a long time,
thereby improving long-term user retention.
Third, we conduct both automatic and manual quality control on the labeled data,
including verifying the existence of citation relationships between \refsources and target papers,
identifying the perfunctory annotations via the quality of the reading notes 
(e.g., incoherent abstract translation without modifications),
and manually checking the rationality of the annotations.

\hide{
\begin{figure*}[t]
	\centering
	\includegraphics[width=\textwidth]{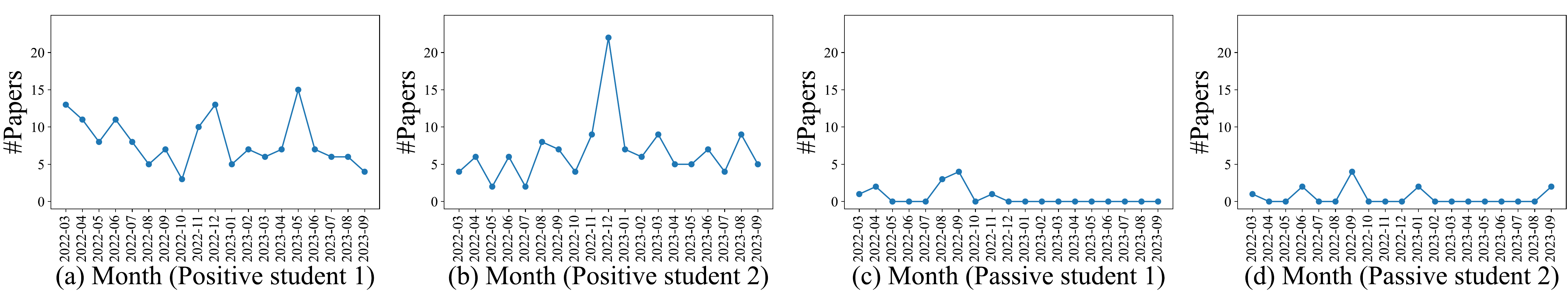}
    \caption{Positive and passive student patterns.}
	\label{fig:user_patterns}
\end{figure*}

\begin{figure}[t]
	\centering
	\includegraphics[width=6cm]{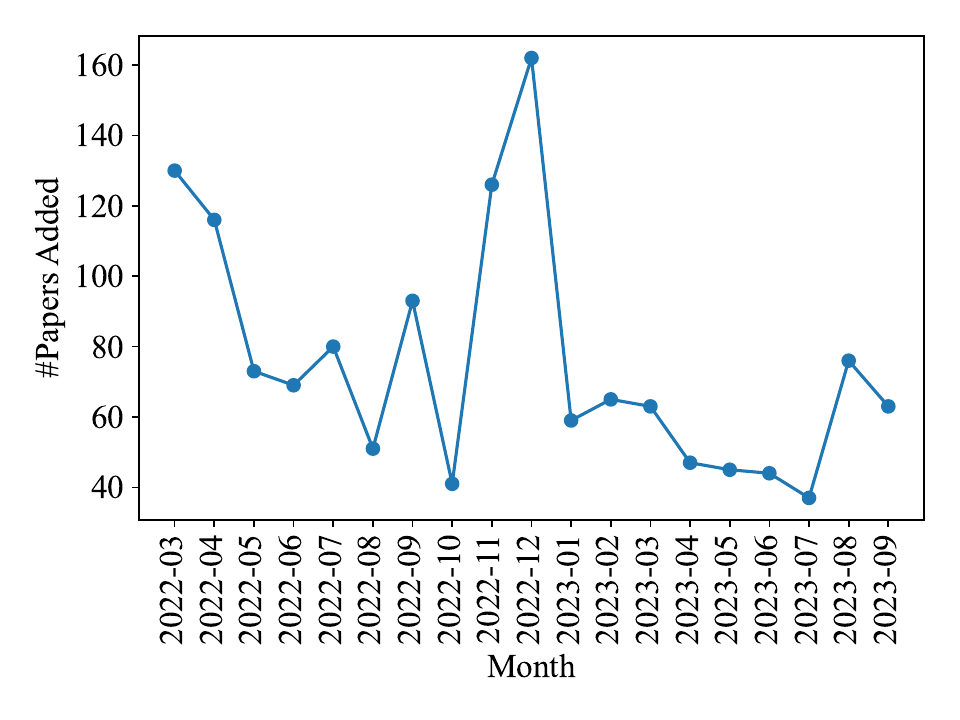}
    \caption{New papers added per month in the paper reading group.}
	\label{fig:monthly_add_cnt}
\end{figure}
}

\vpara{Human evaluation.}
Senior researchers double-checked $100$ papers in the test set 
and tried to identify those papers that were clearly annotated incorrectly.
The sampled correct rate is $94\%$.
\section{Preliminary Study}
\label{sec:analysis}

\subsection{Overall Analysis of \pstbench}

\begin{figure}[t]
	\centering
	\includegraphics[width=7cm]{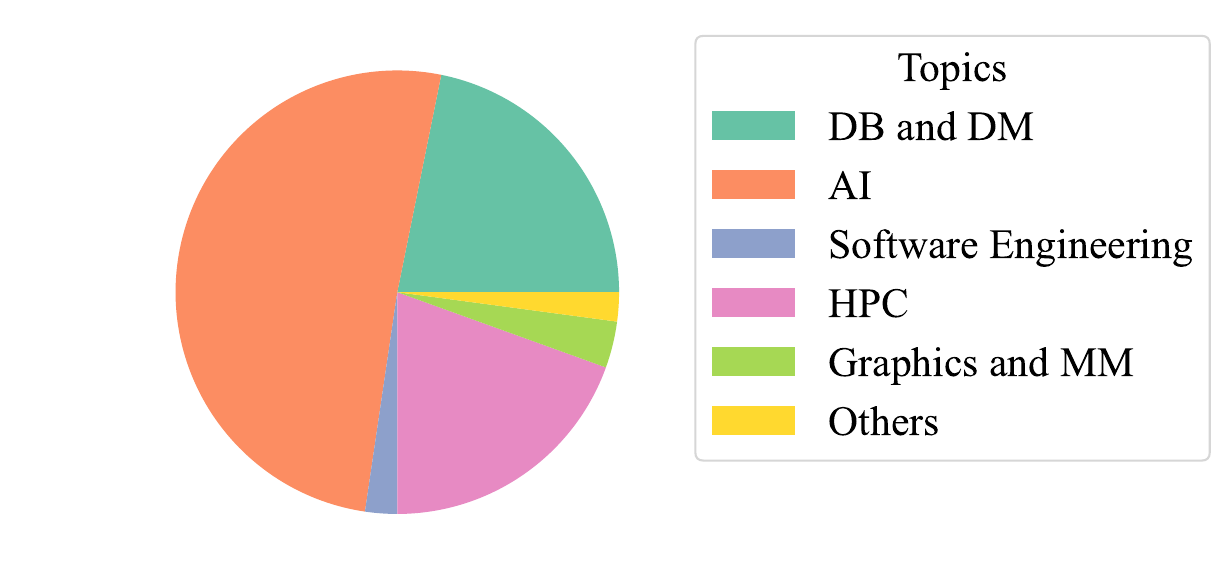}
    \caption{Paper topic distribution.
	\textmd{\small DB and DM: Database and Data Mining,
	AI: Artificial Intelligence and Pattern Recognition,
	HPC: High Performance Computing,
	Graphics and MM: Computer Graphics and Multimedia.
	}}
	\label{fig:paper_topic}
\end{figure}

\begin{figure*}[t]
	\centering
	\hspace{-0.05in}
	\mbox{
		
		\subfigure[PST graph]{
			\includegraphics[width=0.5\textwidth]{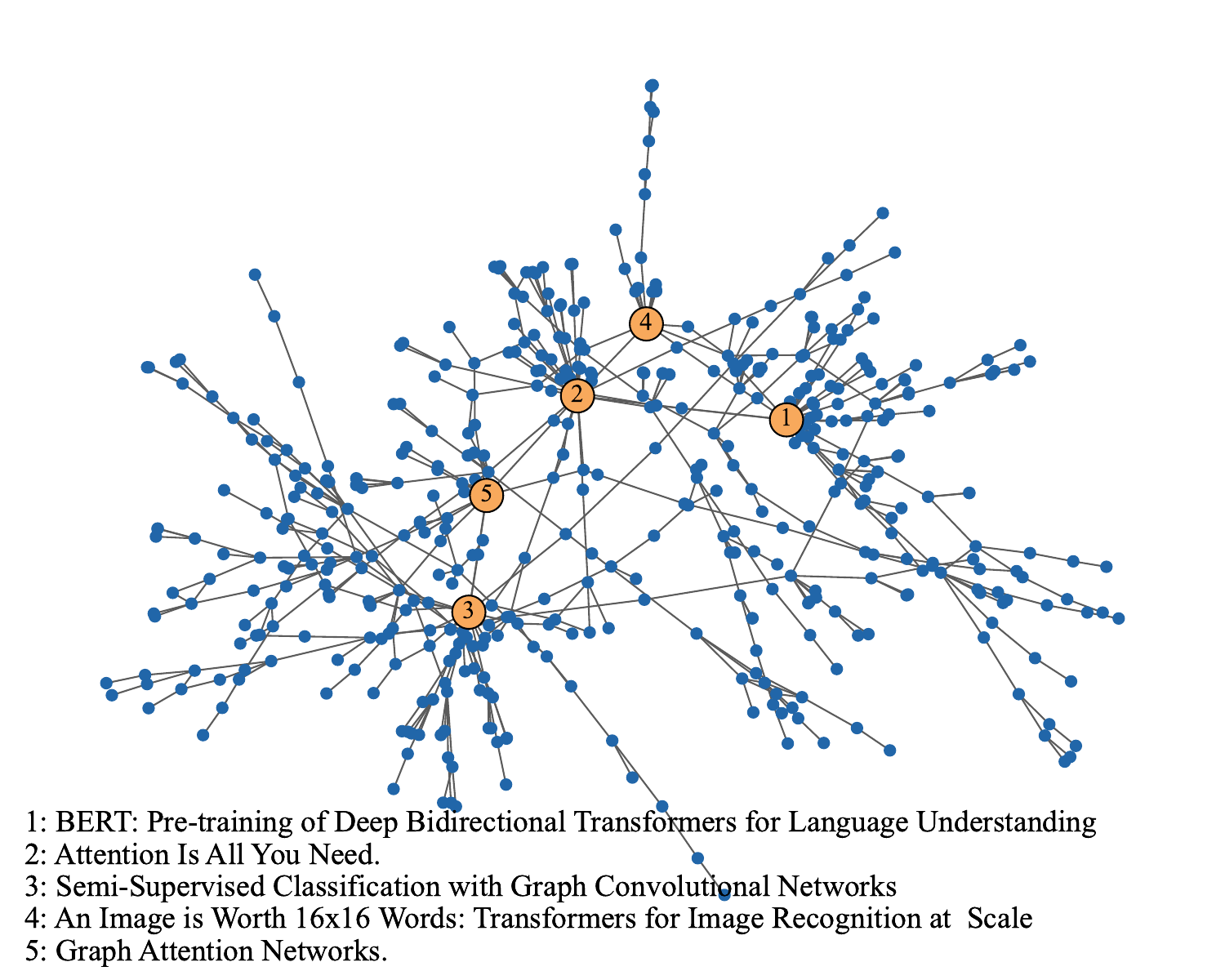}	
			\label{fig:pst_graph}
		}
		
		\subfigure[Citation graph]{
			\includegraphics[width=0.5\textwidth]{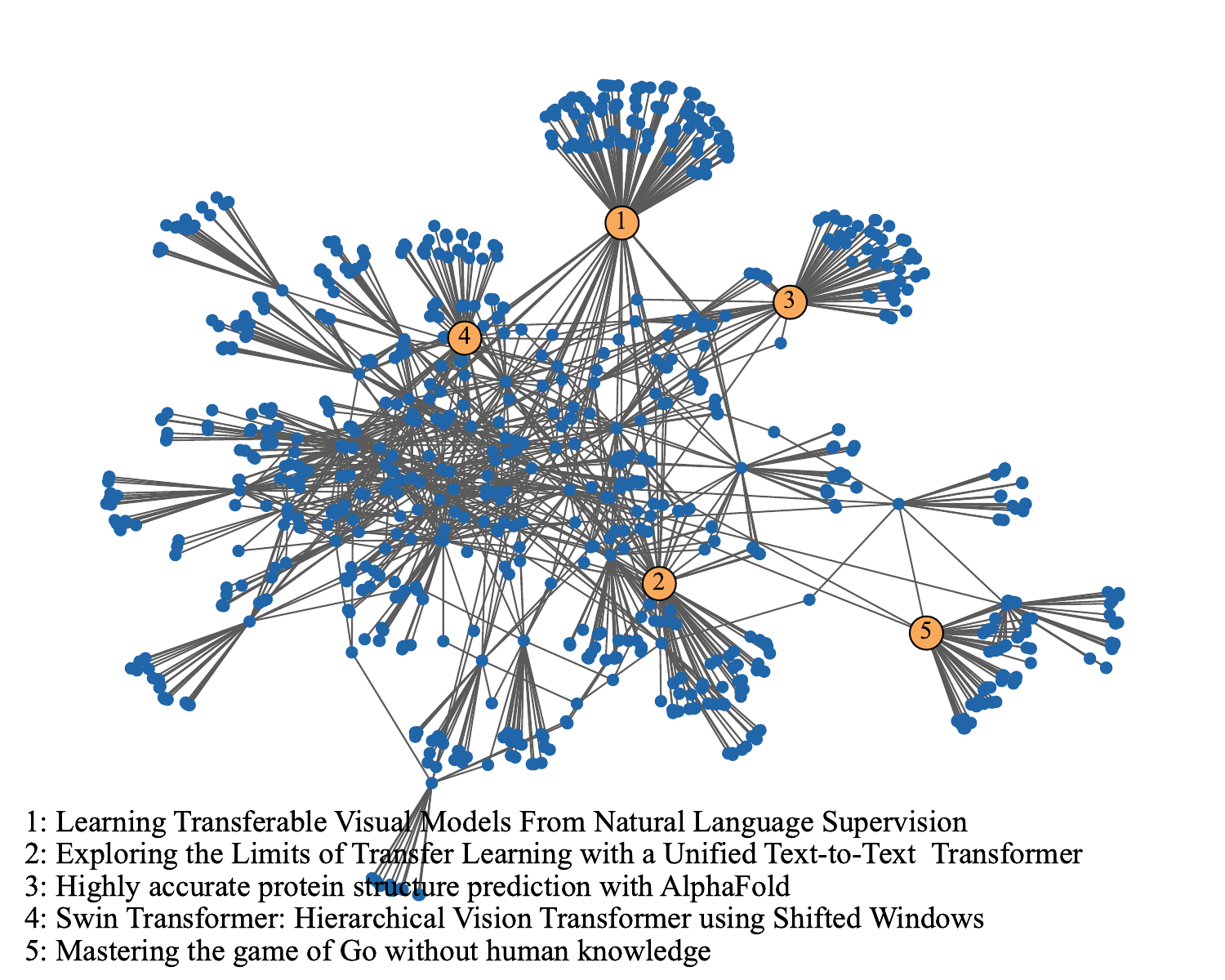}	
			\label{fig:citation_graph}
		}
		
	}
	\vspace{-1.5pt}
	\caption{
		Visualization of the simplified PST graph and the simplified citation graph.
	}
\label{fig:pst_graph_citation_graph}
\end{figure*}

\begin{figure*}[t]
	\centering
	\hspace{-0.05in}
	\mbox{
		
		\subfigure[Distribution of the number of \refsources per paper.]{
			\includegraphics[width=0.31\textwidth]{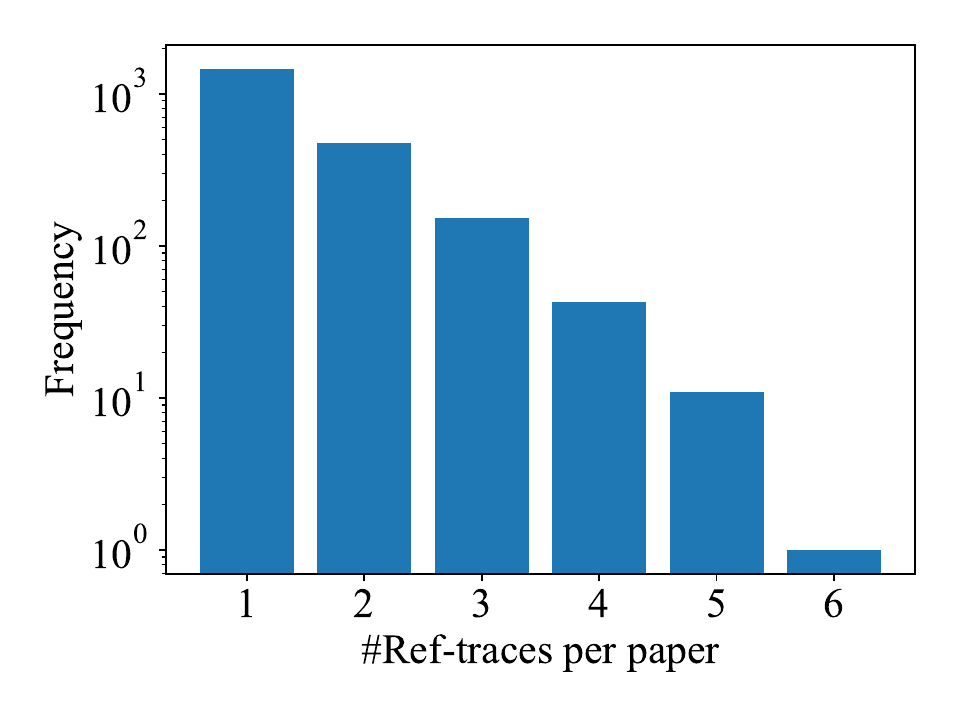}	
			\label{fig:n_ref_traces_freq}
		}
		\hspace{0.1in}
		
		\subfigure[Frequency of a paper being regarded as \refsources.]{
			\includegraphics[width=0.31\textwidth]{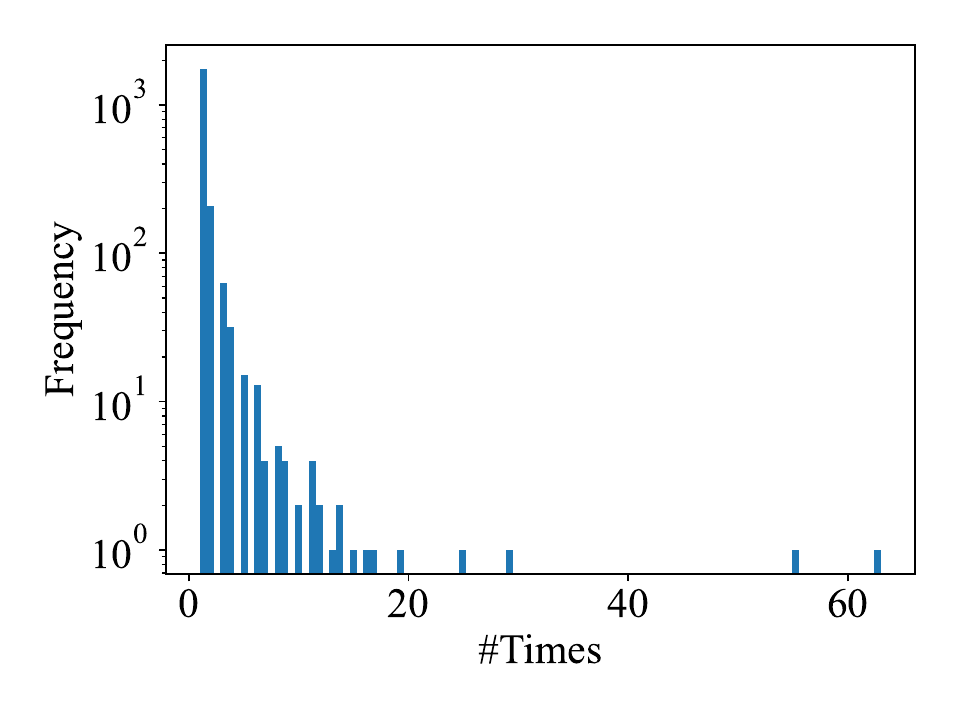}	
			\label{fig:freq_as_ref_traces}
		}
		\hspace{0.1in}
		\subfigure[Cumulative distribution between \refsources and target papers.]{
			\includegraphics[width=0.31\textwidth]{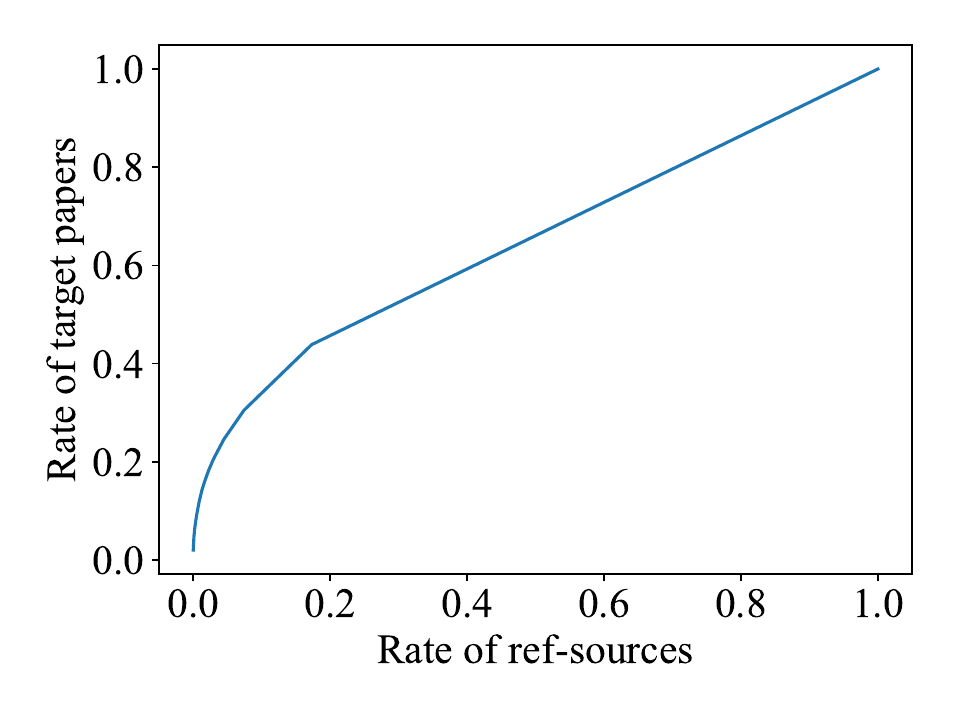}	
			\label{fig:accumulated_freq_ref_sources}
		}
		
	}
	\vspace{-1.5pt}
	\caption{
		Analysis of the distribution of \refsources.
	}
\label{fig:aaa}
\end{figure*}

\hide{
\begin{figure}[t]
	\centering
	\includegraphics[width=8cm]{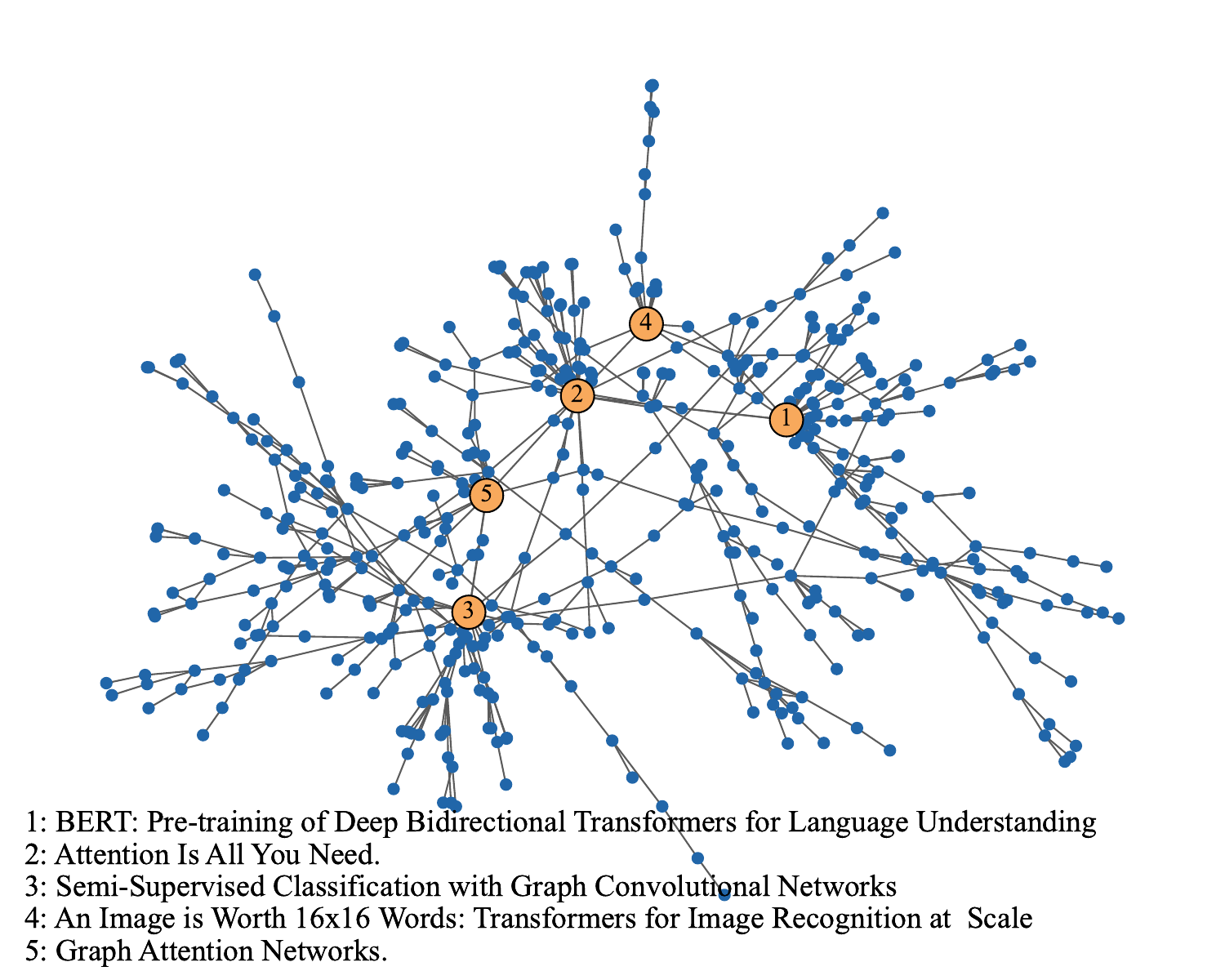}
    \caption{Paper source tracing graph.
    Papers with more than $100$ citations are plotted.
    The edges represent the relations between papers and their \refsources.
    The five nodes with the largest degree are enlarged.}
	\label{fig:paper_citation_graph}
\end{figure}
}

\vpara{Paper topic distribution.}
Figure \ref{fig:paper_topic} visualizes the topic distribution of the collected papers,
which are categorized into five subtopics\footnote{\url{https://numbda.cs.tsinghua.edu.cn/~yuwj/TH-CPL.pdf}}.
This figure reveals that the majority of papers fall within the AI field,
followed by \textit{database and data mining} and \textit{high performance computing (HPC)}.
This distribution is largely due to the fact that our paper reading group initially expanded from students in the HPC and AI groups.
Papers in other fields can be added to the dataset in a similar way in the future.

\begin{figure*}[t]
	\centering
	\hspace{-0.05in}
	\mbox{
		
		\subfigure[AI]{
			\includegraphics[width=4.1cm]{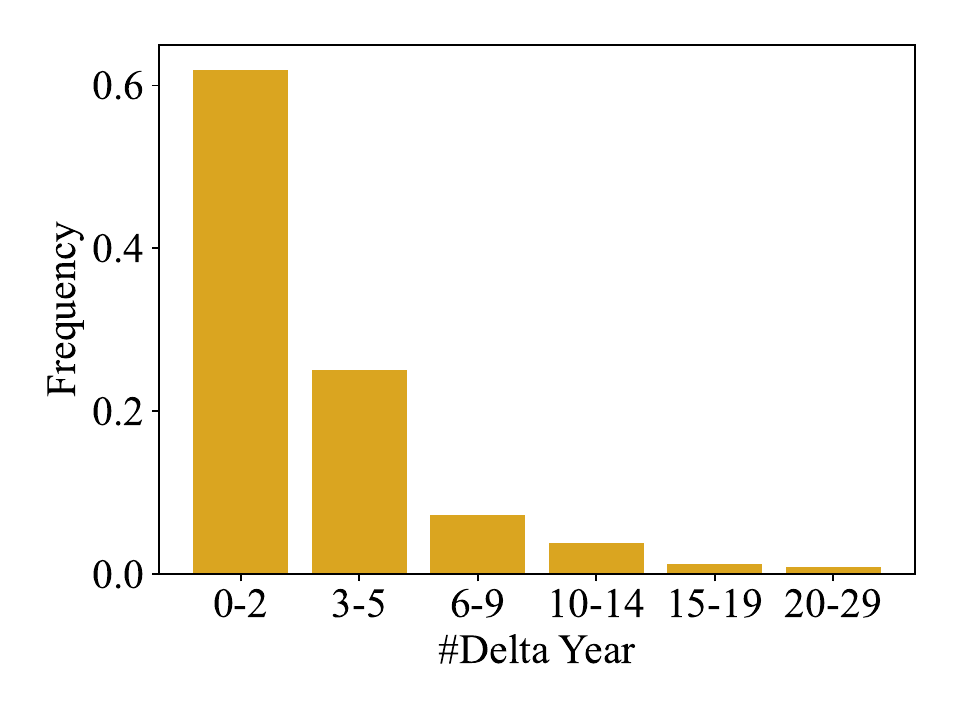}	
			\label{figsub:delta_year_ai}
		}
		\hspace{-0.2in}
		
		\subfigure[Computer Graphics]{
			\includegraphics[width=4.1cm]{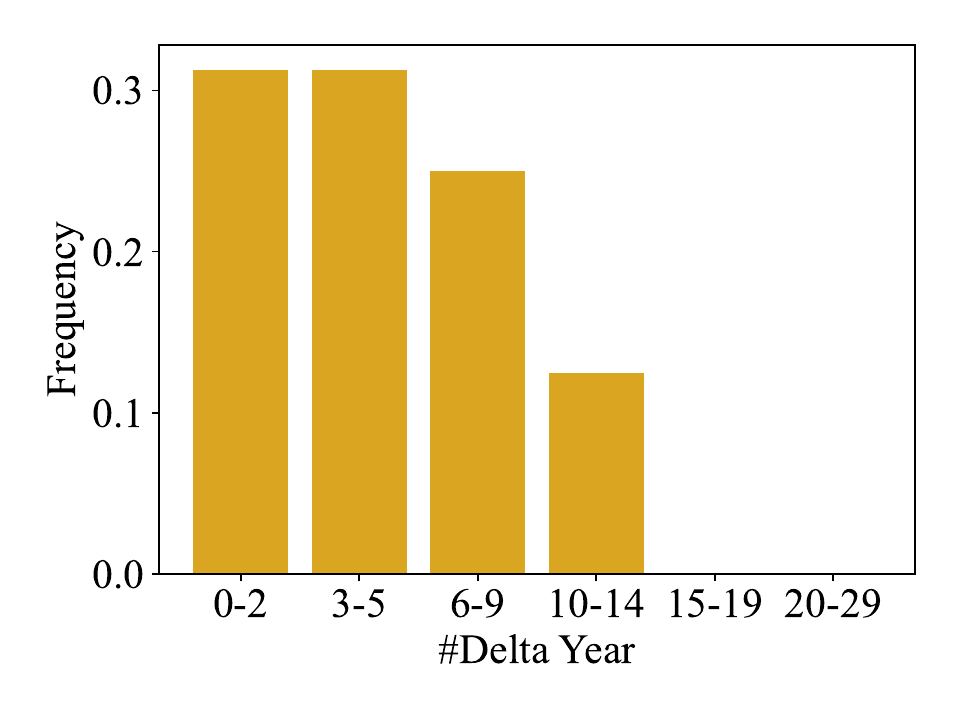}	
			\label{figsub:delta_year_cg}
		}
		\hspace{-0.2in}
		\subfigure[Database and Data Mining]{
			\includegraphics[width=4.1cm]{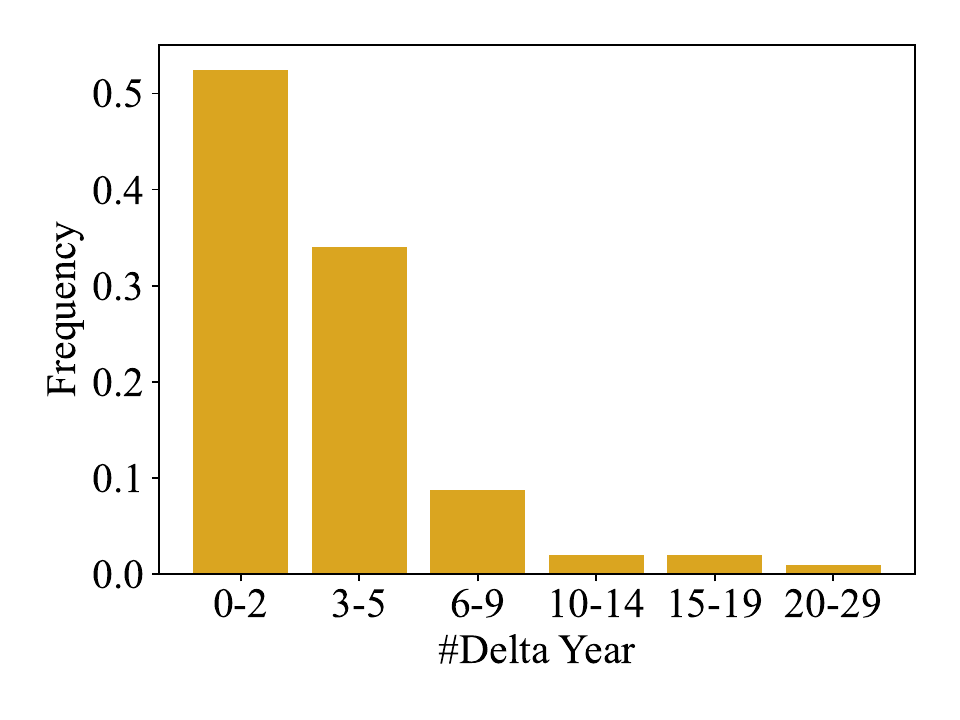}	
			\label{figsub:delta_year_db}
		}
		
		\hspace{-0.2in}
		\subfigure[HPC]{
			\includegraphics[width=4.1cm]{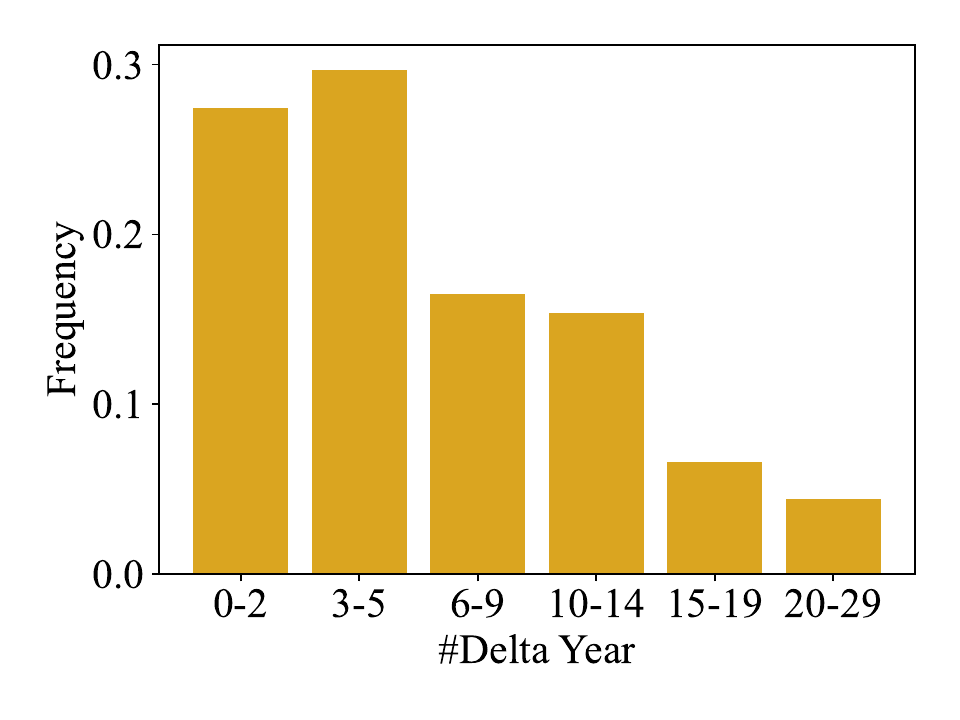}	
			\label{figsub:deltayear_hpc}
		}
		
	}
	\vspace{-1.5pt}
	\caption{
		Year gap between a paper and its \refsources in different fields.
	}
\label{fig:delta_year_all}
\end{figure*}

\begin{figure*}[t]
	\centering
	\hspace{-0.05in}
	\mbox{
		
		\subfigure[AI]{
			\includegraphics[width=0.37\textwidth]{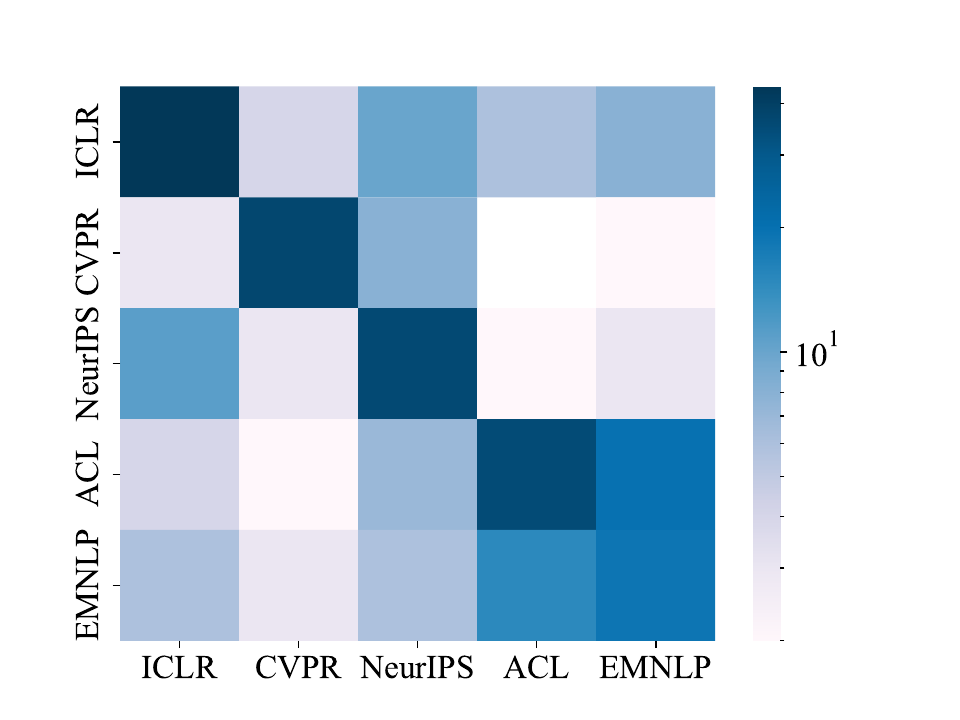}	
			\label{fig:heatmap_v2v_ai}
		}
		\hspace{-0.3in}
		
		\subfigure[Database and Data Mining]{
			\includegraphics[width=0.37\textwidth]{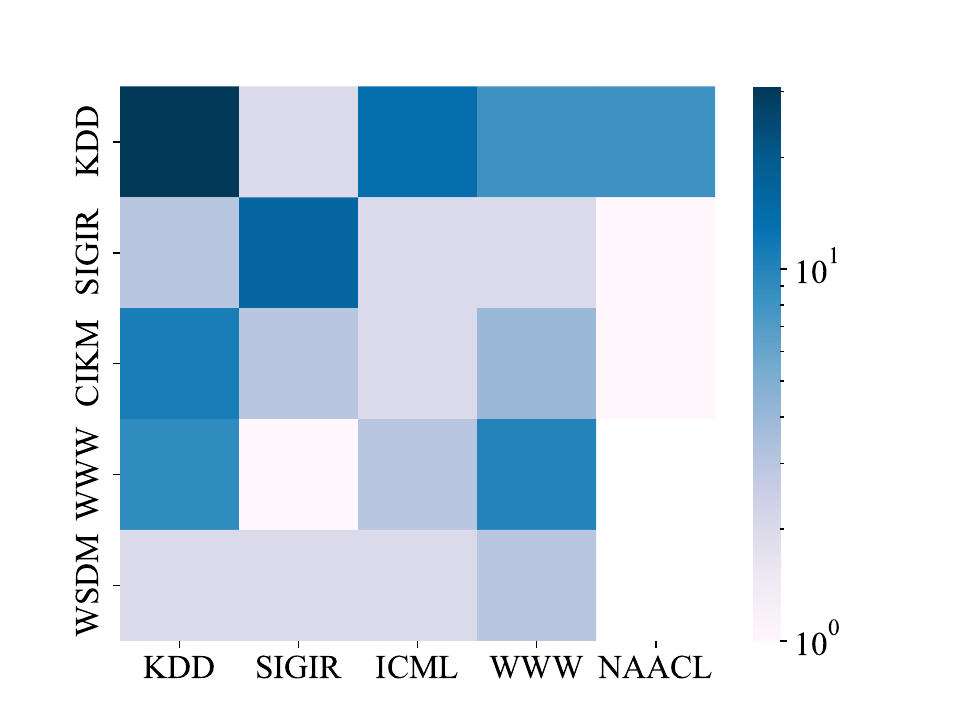}	
			\label{fig:heatmap_v2v_dm}
		}
		\hspace{-0.3in}
		\subfigure[HPC]{
			\includegraphics[width=0.37\textwidth]{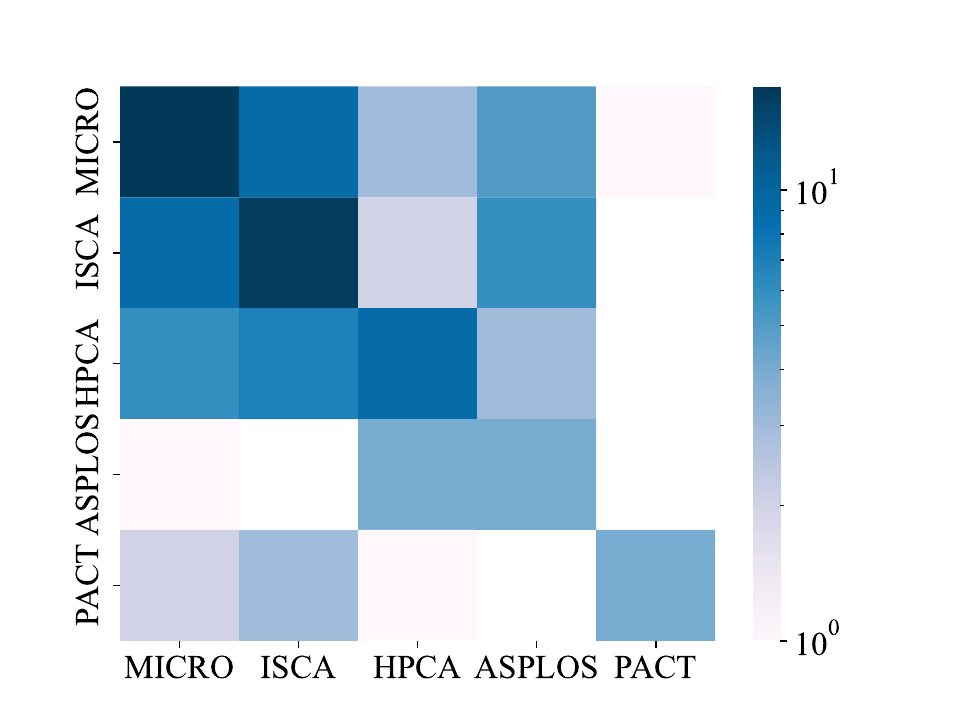}	
			\label{fig:heatmap_v2v_hpc}
		}
		
	}
	\vspace{-1.5pt}
	\caption{
		Influence between computer science venues.
	}
\label{fig:heatmap_v2v_all}
\end{figure*}

\vpara{PST graph vs. citation graph.}
The PST graph, denoted as $\mathcal{G}_{\text{pst}} = \{\mathcal{P}, \mathcal{E}\}$,
consists of a paper set $\mathcal{P}$
and edge set $\mathcal{E}$.
Each edge $e \in \mathcal{E}$ represents the relations between
one paper and its \textit{ref-sources}.
For better visualization,
we plot the largest connected component of the PST graph, 
including paper nodes with over $100$ citations, 
in Figure \ref{fig:pst_graph}.
We discover that papers are scattered in several ``communities'',
each containing a ``super node''.
This figure vividly illustrates the research threads of several fields. 
For instance, 
Transformers~\cite{vaswani2017attention} (\textit{node 2}) 
and BERT~\cite{devlin2019bert} (\textit{node 1})
inspired a significant body of pre-training works,
including ViT~\cite{dosovitskiy2020image} (\textit{node 4}).
ViT, in turn, inspired numerous research works in computer vision.
On the left,
graph convolutional networks (GCN)~\cite{kipf2016semi} (\textit{node 3})
and its subsequent inspired graph attention networks (GAT)~\cite{velickovic2017graph} (\textit{node 5})
are two pioneering works
that inspired a lot of studies in graph mining.

Additionally, we plot the corresponding citation graph in Figure \ref{fig:citation_graph} for comparison.
Given the density of citations, 
we visualize paper nodes with at least \num{10000} citations.
Despite this simplification, 
Figure \ref{fig:citation_graph} is much denser than Figure \ref{fig:pst_graph}.
Figure \ref{fig:citation_graph} presents more diverse research fields,
including language-image pretraining, natural language pretraining,
protein pretraining, vision pretraining, etc.
In Figure \ref{fig:citation_graph},
Swin Transformer~\cite{liu2021swin} (\textit{node 4})
cites CLIP~\cite{radford2021learning} (\textit{node 1}),
and CLIP cites T5~\cite{raffel2020exploring} (\textit{node 2}).
However, 
these citation relationships exist primarily due to background introductions
and don't represent the evolution of relevant fields.
Thus, it is arduous to identify the evolution of these research works from the intricate citation graph.

\subsection{Distribution Analysis of \refsources}

In the following subsection, 
we conduct a detailed analysis of the distribution of \refsources. 

\vpara{\textit{Ref-sources} per paper.}
Figure \ref{fig:n_ref_traces_freq} depicts the histogram of the number of \refsources per paper.
It demonstrates that most annotated papers have only one \refsource,
with about $10\%$ of papers having more than three \refsources.
This could reflect the actual distribution of \refsources per paper to some extent,
suggesting that the majority of annotated papers are inspired by one significant idea.

\vpara{Matthew effect of \refsources.}
Figure \ref{fig:freq_as_ref_traces} and Figure \ref{fig:accumulated_freq_ref_sources}
display the frequency of a paper being considered as a \refsource
and the cumulative distribution between \refsources and target papers, respectively.
We observe that the majority of papers are regarded as \refsources only \textbf{once} in our dataset,
while only a few dozen papers are regarded as \refsources more than $10$ times.
In Figure \ref{fig:accumulated_freq_ref_sources},
the rate of \refsources is sorted by the times of a paper being treated as a \refsource.
We observe that the top $20\%$ of papers inspire more than $40\%$ of other papers,
and the top $40\%$ of papers inspire about $60\%$ of papers.
Papers ranked in the bottom $20\%$ largely maintain a one-to-one mapping with their \refsources,
demonstrating the diversity of related research as well as our datasets.



\subsection{Analysis of Different Topics}

What are the underlying evolution patterns of different topics?
In the following subsection, we conduct analyses from multi-faceted aspects.

\vpara{How soon will one \refsource inspire subsequent works?}
We examine the year gap between a paper and its \refsources across different fields.
Figure \ref{fig:delta_year_all} shows the distribution of the year gap 
in four fields with the most papers.
We have the following intriguing observations.
(1) Across all studied fields, most papers are inspired by \refsources published within the past five years.
Papers are less likely to be influenced by older publications.
(2) Clear differences between fields exist in terms of the distribution of the year gap.
For example, in HPC and computer graphics, 
roughly the same order of magnitude of papers are inspired by 
papers from 0-2 years ago and papers from 3-5 years ago.
However, in AI and \textit{database and data mining},
almost half of the papers are inspired by papers from 0-2 years ago.
Some HPC papers are even inspired by papers published more than $20$ years ago,
a phenomenon rarely seen in other fields.
It reveals that some areas, such as AI, are developing rapidly,
while for fields such as HPC, 
papers in these fields tend to have a relatively longer life force.

\vpara{Influence between computer science venues.}
For target venues in each subtopic,
we study \refsources in which source venues are more likely to 
inspire papers in target venues.
We count pairwise influence relationships between venues,
selecting the subtopics with the most annotated papers,
including AI, database and data mining, and HPC.
For each subtopic, we select the top-$5$ target venues with the most papers 
and top-$5$ source venues that inspired most papers in target venues.
Figure \ref{fig:heatmap_v2v_all} displays the heatmaps of pairwise venue influence on these subtopics.
We highlight several observations below.
(1) AI venues are mostly influenced by AI venues.
(2) In addition to being affected by data mining (DM) conferences,
DM conferences are also influenced by AI conferences (e.g., ICML and NAACL).
(3) HPC conferences are primarily influenced by HPC conferences.
These figures clearly demonstrate the cross-influence between different fields in computer science.

\begin{figure}[t]
	\centering
	\includegraphics[width=6cm]{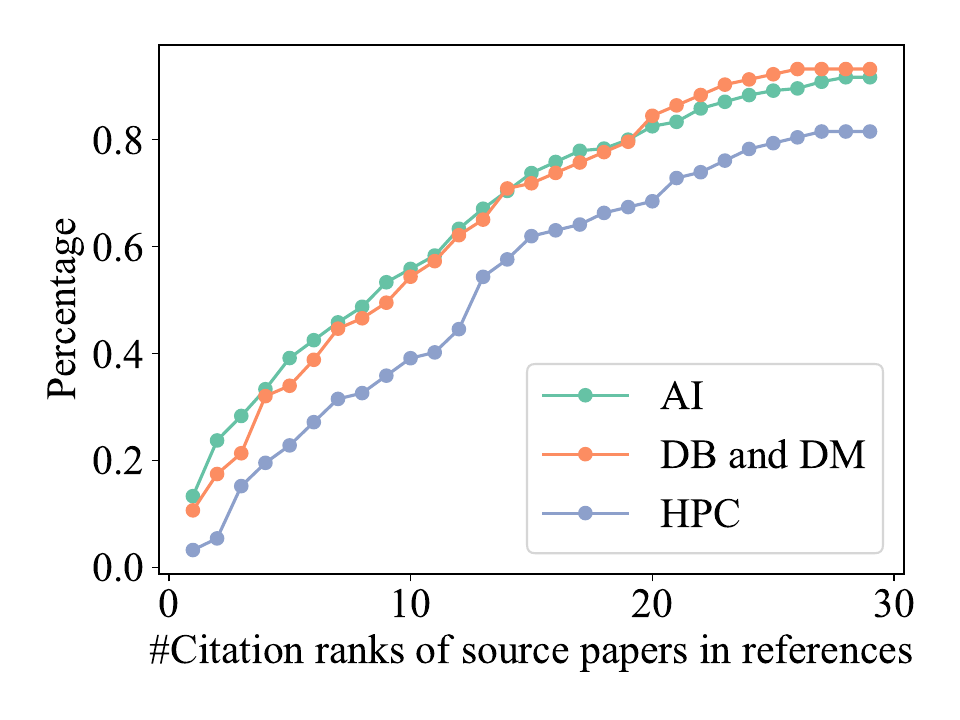}
    \caption{The cumulative distribution function (CDF) 
	of the references' citation ranks for \refsources
	with respect to different topics.}
	\label{fig:cdf_citation_rank_topics}
\end{figure}

\vpara{Are papers inclined to be inspired by the most cited references?}
Figure \ref{fig:cdf_citation_rank_topics} plots the 
CDF of the references’ citation ranks for being \refsources w.r.t. different topics.
The HPC curve is clearly below the curves of AI, DB and DM.
That is, compared with the other two fields,
despite the fact that HPC papers tend to be motivated by older publications,
it doesn't mean that HPC papers are more likely to be inspired by most cited references.
The correlation between the citation number of a reference
and its probability of being a \refsource is weaker in HPC than in AI, DB and DM.
\section{PST Approach}

With the vast proliferation of research papers,
manually annotating the source of each paper is impractical.
Can we automatically identify the \textit{ref-sources} of a paper?
In this section, we explore various approaches to address the PST problem.
PST approaches can be broadly categorized into the following classes:
(1) statistical methods,
(2) graph-based methods,
and (3) pre-trained language model (PLM) based methods.

\subsection{Statistical Methods}

\vpara{Rule.}
An intuitive method to discover \textit{ref-sources} is the rule-based method,
which extracts references that appear near signal words like ``motivated by'' or ``inspired by''.
Nevertheless, a limitation of this method is that 
not all \textit{ref-sources} are explicitly mentioned in proximity to these signal words.

\vpara{Random Forest (RF).}
Alternatively, we can define statistical features
related to each 
reference
to indicate its importance.
Following~\cite{valenzuela2015identifying},
we define features including citing count, citing position,
author overlap, text similarity, etc.
We then employ RF to classify the importance of each reference.
RF is adopted due to its effectiveness in filtering out unrelated features.

\subsection{Graph-based Methods}

The paper citation graph can also deliver the structural importance or structural similarity of each reference to the target paper.
For instance, an extension paper $p_e$ and its original paper $p$ probably share many references.
Thus, their structural similarity should be high.
To this end, we extract the paper citation graph 
in computer science\footnote{\url{https://www.aminer.cn/citation}} 
and learn paper embeddings with network embedding methods, such as
\textbf{LINE}~\cite{tang2015line}, \textbf{ProNE}~\cite{zhang2019prone}, \textbf{NetSMF}~\cite{qiu2019netsmf}.
We adopt these methods owing to their effectiveness and efficiency in handling large-scale graphs.
Next, we measure the importance of references to the target paper 
by calculating the cosine similarity between the paper representation and the reference representation.

\subsection{PLM-based Methods}

Imagine how researchers judge whether a reference is a \refsource.
They may read the context where the reference appears in the full text of the paper 
and then decide whether the reference is a \refsource based on content comprehension.
Recently, pre-trained language models (PLMs) have achieved great success in various natural language understanding tasks.
Hence, we can extract the contextual texts where each reference appears in the full text 
and then encode these texts with the pre-trained models,
which are then 
followed by an MLP classifier for binary prediction.
We use the annotation results in the training set as supervision information
to fine-tune the parameters of pre-trained models and the classifier layers.
Then, fine-tuned models are used to predict the \refsources of papers in the test set.
The considered PLMs include
\textbf{BERT}~\cite{devlin2019bert}, \textbf{SciBERT}~\cite{beltagy2019scibert}, 
\textbf{GLM}~\cite{du2022glm}, and \textbf{Galactica}~\cite{taylor2022galactica}.
We also adopt three state-of-the-art closed-source models:
\textbf{GPT-3.5}~\cite{openai2022chatgpt}, 
\textbf{GPT-4}~\cite{achiam2023gpt}, 
and \textbf{Claude}~\cite{anthropic2023claude}.


\hide{
\subsection{Ensemble Methods}
To leverage the strengths of
each category of methods,
we employ an ensemble method to combine the predictions of different methods.
Specifically, we select the best performer from each category of methods 
and average their predictions as the final prediction.
We opt for average instead of voting to avoid 
specifying thresholds for each method.
}
\section{Experiments}
\label{sec:exp}

\subsection{Experimental Setup}
\label{sec:exp:setup}

For the full texts of papers,
we use the GROBID\footnote{\url{https://grobid.readthedocs.io/en/latest/}} API to 
convert PDF to XML format for convenient processing of citation contexts.
We employ regular expressions to identify the contexts of each reference.
For graph-based methods, the node embedding size is set to $128$.
We utilize the CogDL~\cite{cen2023cogdl} framework to implement graph-based methods.
For PLM-based methods,
the context length is set to $200$.
More implementation details can be found in Section \ref{sec:app:implementation}.

\begin{table}[t]
    \centering
    \caption{Accuracy results of paper source tracing. 
    }
    \begin{threeparttable}
    \renewcommand\tabcolsep{10pt}
    \renewcommand\arraystretch{1.05}
    \begin{tabular}{c|c|c}
        \toprule[1.2pt]
            & Method &   MAP      \\
         \midrule
        \multirow{2}{*}{Stat} 
        & Rule     &  0.0616     \\
        & RF     &  0.1821   \\
        \midrule
        \multirow{3}{*}{Graph} 
        & LINE  & 0.1047 \\
        & ProNE     &  0.1050   \\
        & NetSMF & 0.1231 \\
         \midrule
        \multirow{7}{*}{PLM}
        & BERT-base & 0.2775 \\
        & SciBERT & \textbf{0.3240} \\
         & GLM-2B  &  0.1503  \\
         & Galactica-standard & 0.1472 \\
         & GPT-3.5 & 0.0781 \\
         & GPT-4 & 0.0519 \\
         & Claude-instant & 0.0536 \\
        \bottomrule[1.2pt]
    \end{tabular}
    \begin{tablenotes}
        \footnotesize
        \item Stat: statistical methods.
    \end{tablenotes}
    \end{threeparttable}
    \label{tab:pst_main_results}
\end{table}

\vpara{Evaluation Metrics.}
We adopt mean average precision (MAP) to evaluate the prediction results.
Concretely, for each paper $p$ in the test set,

\beq{
    \text{AP}(p) = \frac{1}{R_p}\sum_{k=1}^{M_p} \text{Prec}_p(k) \mathbbm{1}_k,
}

\noindent where $R_p$ is the number of \refsources of paper $p$,
$M_p$ is the number of references of paper $p$,
$\text{Prec}_p(k)$ is the precision at cut-off $k$ in the ranked output list $S_p(k)$,
and $\mathbbm{1}_k$ is the actual annotation,
with the values $0$ or $1$.

\beq{
    \text{MAP} = \frac{1}{|\mathcal{P}_{\text{test}}|} \sum_{p \in \mathcal{P}_{\text{test}}} \text{AP} (p),
}

\noindent where $\mathcal{P}_{\text{test}}$ is the paper set in the testing set.

\subsection{Main Results}

Table \ref{tab:pst_main_results} presents the results of paper source tracing. 
Among statistical methods,
Random Forest (RF) surpasses the Rule method, 
emphasizing the efficacy of feature engineering.
The Rule-based approach underperforms, 
likely due to the absence of signal words such as ``inspired by'' around many crucial references, 
leading to a low recall rate. 

In terms of graph-based methods,
NetSMF outperforms LINE and ProNE,
likely due to 
its ability to capture higher-order proximity of nodes via sparse matrix factorization.

As for PLM-based methods,
SciBERT significantly surpasses other models,
demonstrating the effectiveness of pre-training on domain-specific data.
Surprisingly, finetuned SciBERT and BERT-base surpass larger models like GLM-2B, Galactica-standard,
and closed-source PLMs.
The reason may lie in two aspects.
First, the training objective of the masked language model is more suitable for this context understanding task.
Second, API-based models may not be adequately trained on similar tasks.
However, the results of current methods are not yet optimal,
suggesting significant potential for further research in this field,
such as combining multiple categories of methods.

\begin{table}[t]
    \centering
    \caption{The feature contribution analysis for RF.}
    {
    \begin{threeparttable}
    \renewcommand{\arraystretch}{1}%
        {
            \setlength{\extrarowheight}{1pt}
            \begin{tabular}{
                    @{}l@{ } c@{}}
                \noalign{ \hrule height 1pt}
                \textbf{\textbf{Feature description}}   & \textbf{Weight} \\ \hline
                citation number of the reference & 0.48 \\
                reciprocal of the number of references & 0.26 \\
                number of paper citations / all citations$^1$ & 0.17 \\
                appearing near signal words$^2$ & 0.02 \\
                author overlap$^3$ & 0.02 \\


             
                \noalign{\hrule height 1pt}
            \end{tabular}}
            \begin{tablenotes}
                \footnotesize
                \item[1] This feature computes the number of direct citation instances for the cited paper over all the direct citation instances in the citing work.
                \item[2] Signal words include ``inspired by'' and ``motivated by''.
                \item[3] Set to true if the citing and the cited works share at least one common author.
            \end{tablenotes}
        \end{threeparttable} 
        }
        \label{tb:feat_contribution}
\end{table} 

\subsection{Feature Analysis}

We conduct a feature importance analysis for random forest,
with the most significant features 
shown in Table \ref{tb:feat_contribution}.
We observe that the most important feature is the citation number of the reference,
aligning
with our previous analysis.
In addition, the number of direct citations of a reference also matters,
which makes sense as the more times a reference is cited, the more important it might be.
Surprisingly, the feature of appearing near signal words is not that important,
possibly due to the sparsity of this feature.
Author overlap is weakly positively correlated with being a \refsource,
which is intuitive since some authors are likely to extend the ideas or methods from their previous works.

\subsection{Error Analysis}

\begin{figure}[t]
	\centering
	\includegraphics[width=8cm]{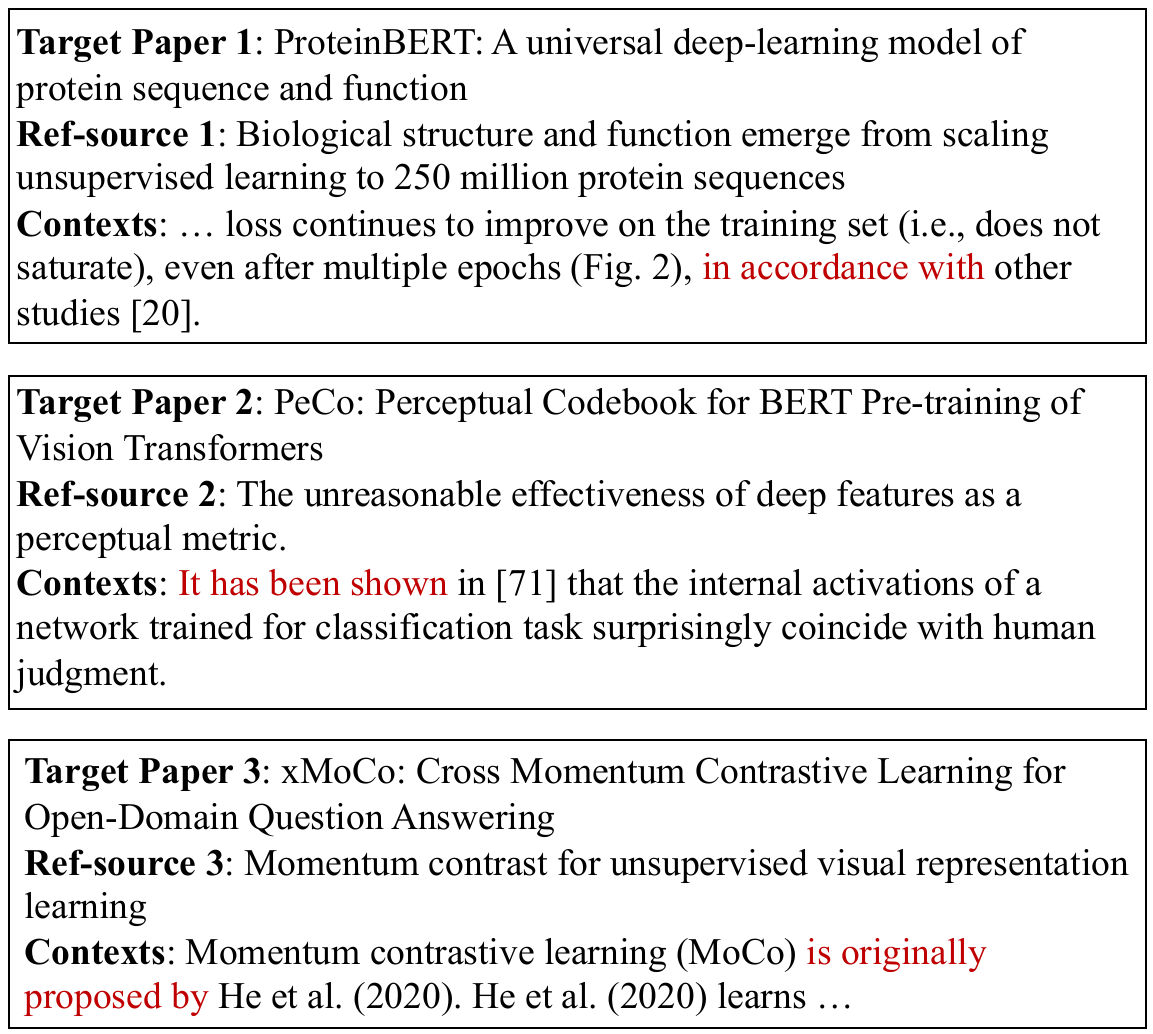}
    \caption{Predictive error analysis.}
	\label{fig:exp:error_analysis}
\end{figure}

We conduct a case study of the prediction errors made by our best-performing model,
with several examples shown in Figure \ref{fig:exp:error_analysis}.
We list each target paper with its \refsource and the corresponding contexts.
We have the following observations.
For target paper $1$, the relationship between the target paper and its \refsource is weak, 
as indicated by the signal words ``in accordance with'',
making it hard to identify the \refsource based on the context.
For target paper $2$, the \refsource appears as a background explanation of the target paper,
resulting in a loose semantic correlation between them.
For target paper $3$, the \refsource is introduced in the related work section 
and is not explicitly associated with the target paper.
However, familiar researchers can easily identify the \refsource 
based on the title similarity of the two papers.
Thus, the general understanding of the main ideas of papers might be omitted in the current contextual methods.
\section{Related Work}

Paper source tracing is closely related to citation intention analysis,
trend analysis, and citation impact evaluation, among others.
The creation of a scalable benchmark dataset that quantifies and annotates the semantics of citation links presents a significant challenge. 
\citet{tang2009topic} 
conduct a study on citation semantic analysis, defining three categories for each citation link: drill down, similar, and others.
They construct a dataset 
comprising approximately $1{\small,}000$ citation pairs in computer science.
Hereafter, \citet{valenzuela2015identifying} propose a new dataset of $450$ citation pairs 
with both incidental and important citations.
\citet{jurgens2018measuring} introduce a larger dataset of nearly $2{\small,}000$
citation pairs in the NLP area,
in which less than $100$ citation pairs are annotated as the motivation.
Most of these datasets 
involve meticulous annotation of each paper, 
comparing one target paper with each reference,
thus making them hard to scale up.

Some endeavors have been made to automatically identify the importance of references. 
Early attempts define hand-crafted features and 
then employ classifiers to determine the significance of references.
\citet{pride2017incidental} argue that abstract similarity is 
one of the most predictive features.
\citet{hassan2017identifying} incorporate several new features,
such as context-based and cue words-based features,
and utilize Random Forest to assess the importance of references.
\citet{he2009detecting} adapt the  LDA 
model to citation networks
and develop a new inheritance topic model to depict the topic evolution.
\citet{farber2018cite} present a convolutional recurrent neural network based method
to classify potential citation contexts.
\citet{jiang2023contextualised} propose contextualized representation models based on
SciBERT~\cite{beltagy2019scibert} to classify citation intentions.
The predictive performance is optimistic on certain datasets,
achieving over $90\%$ AUC.

Paper source tracing has numerous practical applications,
including understanding the evolution of a subfield~\cite{shao2022tracing}
and assessing scholarly impact.
Several online systems,
such as MRT~\cite{yin2021mrt} and IdeaReader~\cite{li2022ideareader},
have been developed to assist researchers in better 
understanding the evolution of ideas or concepts.
Characterizing important references enables a better evaluation of scholarly impact.
\citet{manchanda2021evaluating} propose CCI,
a content-aware citation impact measure, to quantify the scholarly impact of a publication.

In this study, we build an accurate and scalable benchmark \pstbench
for paper source tracing
and investigate a variety of methods for automatic source tracing.
Extensive experiments underscore the complexity  of the task,
which deserves more in-depth exploration in the future.

\section{Conclusion}

In this paper, we present \pstbench,
a novel, professionally annotated, and ever-growing benchmark for paper source tracing.
We conduct detailed analyses on \pstbench
and offer several insights, 
such as the differing evolution patterns of papers across different topics.
\pstbench facilitates further analysis of the evolution of science
and a deep understanding of the crux of research works, and so on.
We plan to expand the coverage of \pstbench to more topics 
and design elaborate methods to improve the accuracy of the PST problem.

\newpage

\section{Ethical Considerations}

For online publications, \pstbench provides publicly available metadata
and very few parsed full-texts of open-access papers for research purposes.
For data annotation, all annotators gave their informed consent for inclusion before they participated in this study. 

\section{Limitations}
\label{sec:limitation}

While \pstbench provides an accurate and scalable benchmark for paper source tracing,
its current format has the following limitations.
(1) The topics covered in \pstbench are not even, 
with most topics related to AI, data mining, and high performance computing.
In the future, we plan to call for students majoring in different areas to
expand the coverage of \pstbench.
(2) Although we explore various types of methods for automatic paper source tracing,
more advanced methods tailored for the PST problem are absent.
However, the elaborate method design for the PST problem is not the main focus of this paper.
We plan to optimize the methods and call for contributions 
to improve the performance of automatic paper source tracing.

\hide{
Second, annotating the source of papers is subjective to some degree.
Different readers may hold different views on selecting \refsources for the same paper.
This might be alleviated by cross-checking from different readers,
but sometimes identifying the source of a paper may be an open question with no standard answer.
Third, annotators might tend to annotate fewer \refsources than actual ones,
which is deferred to future work by cross-checking from multiple annotators. 
}

\section{Broader Impact}
\pstbench can be used by various communities, 
such as NLP, graph mining, science of science, etc.
One can use them to mine and understand the evolution of science
or develop automatic methods to trace the source of papers, etc.



\bibliography{ref}

\appendix



\section{Data Collection}
\label{sec:app:data_collection}

\begin{figure}[t]
	\centering
	\includegraphics[width=8cm]{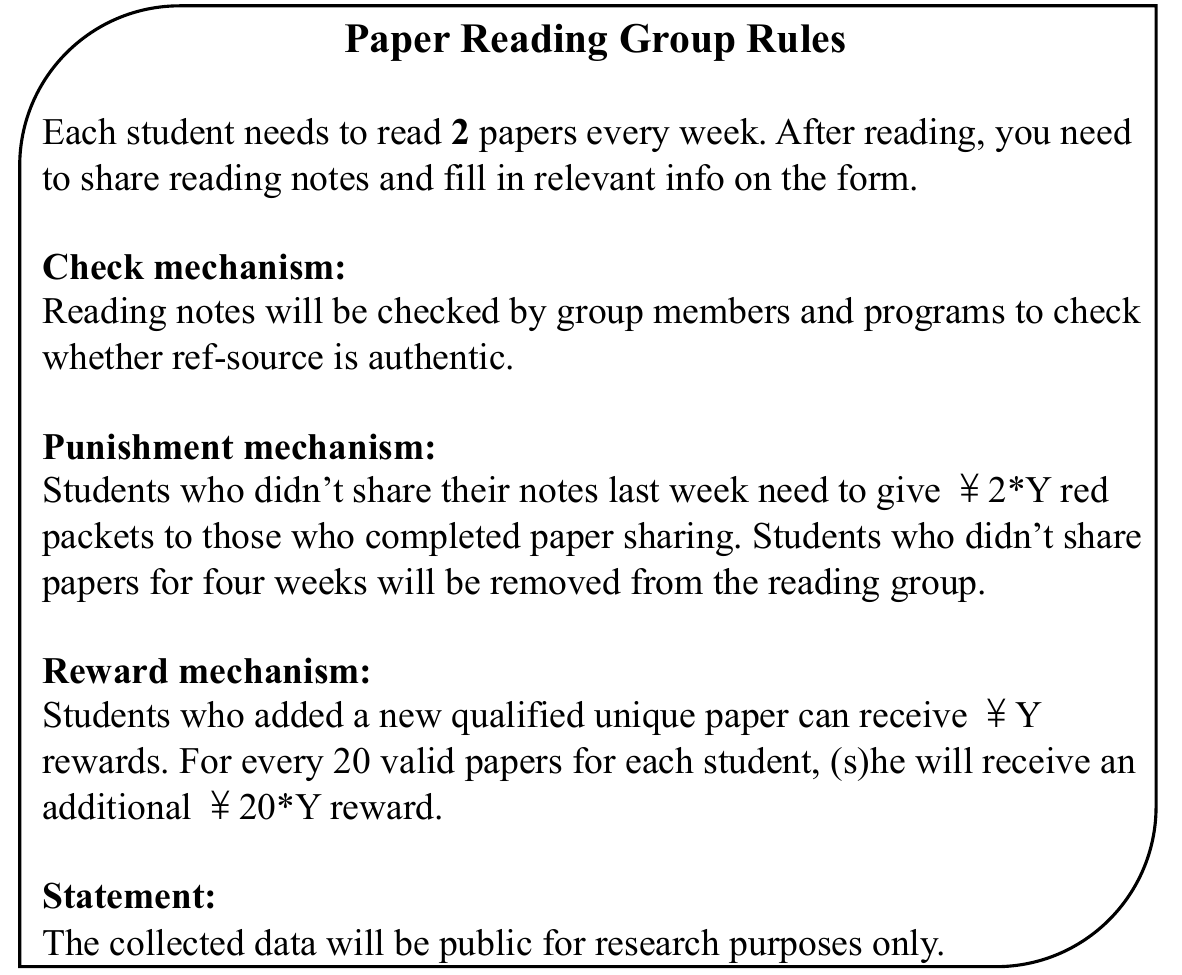}
    \caption{Reading group rules.}
	\label{fig:reading_group_rules}
\end{figure}

The detailed paper reading group rules are shown in Figure~\ref{fig:reading_group_rules}.
Currently, each paper is annotated by one student.
Recruited group members are told that the collected data will be public and used for research purposes only.
Next, we detail the components of data annotation to ensure data quality.

\vpara{Maintenance of the paper reading group.}
We periodically hold paper reading groups on WeChat every week and publicize the reading group 
on the public forums of several universities and familiar labs.
Inactive group members are removed every four weeks.
We remove group members immediately once they have made perfunctory annotations.

\vpara{Reward and punishment mechanism.}
The reward mechanisms are divided into immediate and long-term rewards. As shown in Figure \ref{fig:reading_group_rules},
students receive rewards each week or once they mark every $20$ papers.
In contrast, students who didn't share their paper reading notes need to give red packets to those who completed paper sharing.
The recruited students usually read papers even without the reading group.
Thus, their workload is primarily to annotate the source of papers they have read
and fill in the form we provide.
In this case, the payment is relatively reasonable.

\vpara{Demographics of group members.}
Until February 2024, 
there have been $101$ members 
who participated in the effective annotation.
We don't know many demographics of volunteering students,
but most of them are from China,
studying in renowned universities or research institutes,
including Chinese Academy of Sciences, Tsinghua University, Harbin Institute of Technology,
Southeast University, Nankai University, etc.

\pstbench has been made public under the ODC-BY license.
The created dataset and the original data are used
for research purposes only.
We have anonymized the annotators' information.

\section{Implementation Details}
\label{sec:app:implementation}

\begin{table}[t]
    \centering
    \caption{Parameters and running time of main methods.}
    \begin{tabular}{ccc}
        \toprule
        Method & \#Parameters & Running hours \\
        \midrule
        RF & 12 & 0.05 \\
        LINE & 1.47B & 14 \\
        ProNE & 1.47B & 10 \\
        NetSMF & 1.47B & 16\\
        BERT-base & 110M & 2 \\
        SciBERT & 110M & 2 \\
        GLM-2B & 2B & 5 \\
        Galactica & 6.7B & 13 \\
        \bottomrule
    \end{tabular}
    \label{tb:exp:para_running_time}
\end{table}


\hide{
\begin{itemize}[leftmargin=*]
    \item \textbf{RF}: $12$-dimensional features.
    \item \textbf{LINE, ProNE, NetSMF}: $128$-dimensional embeddings and \num{11478633} papers, resulting in $1.47$B parameters.
    \item \textbf{BERT-base}: $110$M parameters.
    \item \textbf{SciBERT}: $110$M parameters.
    \item \textbf{GLM-2B}: $2$B parameters.
    \item \textbf{GLM-10B}: $10$B parameters.
\end{itemize}
}


The parameters and running time of the main methods are listed in Table~\ref{tb:exp:para_running_time}.
All experiments are conducted on a Linux server with 56 Intel(R) Xeon(R) Platinum 8336C CPU,
1.88T RAM, and 8 NVIDIA A100 GPUs,  each with 80GB memory.

For the fine-tuned BERT, SciBERT, and GLM model, we search for the best learning rate in the range of 
$\{1e^{-5}, 3e^{-5}, 1e^{-4}, 3e^{-4}\}$, and the best learning rate is set to $1e^{-4}$
according to the performance on the validation set.
For the Galactica model, we adopt the Xturing\footnote{\url{https://github.com/stochasticai/xTuring/}} framework
and use the default parameters.
As for API-based methods, we use the same input contexts as other open-sourced pre-trained models for a fair comparison.
We have submitted the paper PDF files and asking the GPT/Claude which references are the most significant to inspire the given papers, 
but the responses are basically unreasonable.
For LINE in CogDL,
we set the \texttt{walk\_length} and \texttt{walk\_num} to $5$ and $5$, respectively.
For NetSMF in CogDL, 
we set the \texttt{window\_size} and \texttt{num\_round} to $5$ and $5$, respectively.
For ProNE in CogDL, we use its default parameters.
For graph-based methods, the constructed citation graph includes \num{11478633} nodes and \num{167161322} edges.
For supervised methods,
we keep all positive instances and sample negative instances randomly, 
keeping their ratio at $1: 10$.

\clearpage

\end{document}